\documentclass[12pt]{article}
\usepackage{graphicx}
\usepackage{amsmath,latexsym,epsfig,epsf,rotate,amssymb}
\begin{document}

\title{Test of FSR in the process $e^+e^-\to\pi^+\pi^-\gamma$ at DA$\Phi$NE \footnote{Talk presented by O. Shekhovtsova at EURIDICE Midterm Collaboration Meeting,Frascati, 8-12 February.}}

\author{G.~Pancheri$^{1)\footnote{e-mail: Giulia.Pancheri@lnf.infn.it}}$
, O.~Shekhovtsova$^{1),2)\footnote{e-mail: shekhovtsova@kipt.kharkov.ua}}$,
 G.~Venanzoni$^{1)\footnote{e-mail: Graziano.Venanzoni@lnf.infn.it}}$ 
\\
\\
\emph{$^{1)}$INFN Laboratori Nazionale di Frascati, Frascati (RM) 00044, Italy} \\
\emph{$^{2)}$NSC ``Kharkov Institute for Physics and Technology'',} \\
\emph{Institute for Theoretical Physics, Kharkov 61108, Ukraine }
} 
\date{}
\maketitle

\begin{abstract}
In this paper we consider the possibility to test the FSR model in the reaction $e^+e^-\to\pi^+\pi^-\gamma$ at DA$\Phi$NE. We propose to consider the low $Q^2$ region ($Q^2$ is the invariant mass squared of the di-pion system) to study the different models describing $\gamma^*\to\pi^+\pi^-\gamma$ interaction. As illustration we compare the scalar QED and Resonance Perturbation Theory prediction for the $e^+e^-\to\pi^+\pi^-\gamma$ cross section. We also consider the  contribution coming from the $\phi$ direct decay ($\phi\to\pi^+\pi^-\gamma$). We find the low $Q^2$ region is sensitive to  FSR models.
\end{abstract}

\textbf{1.} Final state radiation (FSR) is the main irreducible background in radiative return measurements of  the hadronic cross section \cite{kloe} which is  important for the  anomalous magnetic moment of the  muon \cite{amu}.   
Besides being of interest as an important background source, this process could be of interest in itself, because a detailed experimental study of FSR allow us to get information about pion-photon interaction at low energies.

Usually the FSR tensor is evaluated in the scalar QED (sQED) model, or more exactly in the combined sQED$*$VMD model,  i.e. the pions are treated as point-like particles and then the total FSR amplitude is multiplied by the pion form factor calculated in the VMD model \cite{sqed}. Additional contributions to the FSR amplitude are possible. In \cite{our} the FSR tensor was estimated in the framework of the Chiral Perturbation  Theory with the explicit inclusion of the vector  and  axial--vector mesons, $\rho_0(770)$ and  $a_1(1260)$,  called Resonance Perturbation Theory (RPT) \cite{Ecker_89}. We apply the results obtained in \cite{our} for the case of the $\phi$-factory DA$\Phi$NE. 

This paper is organized as follows. In Section 2 we briefly repeat the main results of \cite{our}. In Section 3 we introduce the corresponding results into a Monte Carlo (MC) program, based on the EVA generator of the process $e^+e^-\to\pi^+\pi^-\gamma$ \cite{binner}. In Section 4 we present our conclusions. 

\textbf{2.} Based on charge-conjugation symmetry, photon crossing symmetry and gauge invariance the general amplitude for $\gamma^*(Q)\to\pi^+(p_+)\pi^-(p_-)\gamma(k)$, when the final photon is real, can be expressed by three gauge invariant tensors (see Appendix A in \cite{our} and Refs. [23-25] therein)
\begin{eqnarray}\label{fsr}
&&M^{\mu \nu }(Q,k,l)=-ie^{2}(\tau _{1}^{\mu \nu }f_{1}+\tau
_{2}^{\mu \nu }f_{2}+\tau _{3}^{\mu \nu }f_{3})\equiv
-ie^{2}M_{F}^{\mu \nu }(Q,k,l),  \; \; \; \; l=p_+-p_-, \label{fsr_str}\\
&&\tau _{1}^{\mu \nu }=k^{\mu }Q^{\nu }-g^{\mu \nu }k\cdot Q,  \nonumber \\
&&\tau _{2}^{\mu \nu }=k\cdot l(l^{\mu }Q^{\nu }-g^{\mu \nu }k\cdot
l)+l^{\nu }(k^{\mu }k \cdot l-l^{\mu }k \cdot Q),  \nonumber \\
&&\tau _{3}^{\mu \nu }=Q^{2}(g^{\mu \nu }k\cdot l-k^{\mu }l^{\nu
})+Q^{\mu }(l^{\nu }k\cdot Q-Q^{\nu }k\cdot l).
\nonumber
\end{eqnarray} 
We would like to point out that this decomposition is model independent, while  the exact value of the scalar functions $f_{i}$ are determined by the specific  FSR model.

In sQED for the functions $f_i$ we have \cite{sqed}
\begin{equation}
f_1^{sQED}=\frac{2k\cdot Q}{(k\cdot Q)^2-(k\cdot l)^2}, \; \; \; \; f_2^{sQED}=\frac{-2}{(k\cdot Q)^2-(k\cdot l)^2}, \;\; \; \; f_3^{sQED}=0,
\end{equation}
Because of Low's theorem, these equations imply that for $k\to 0$  we have
\begin{equation}\label{funct_0}
\mathrm{lim}_{k\to 0}f_1=\frac{2k\cdot Q F_\pi(Q^2)}{(k\cdot Q)^2-(k\cdot l)^2}, \hspace{1.5em} \mathrm{lim}_{k\to 0}f_2=\frac{-2 F_\pi(Q^2)}{(k\cdot Q)^2-(k\cdot l)^2}, \hspace{1.5em} \mathrm{lim}_{k\to 0}f_3=0,
\end{equation}
where $F_\pi$ is a VMD pion form factor describing $\gamma^*\to\pi^+\pi^-$. Thus for soft photon radiation the FSR tensor is  expressed in the term of one form factor $F_\pi$, but in general we have  three independent form factors describing the FSR process.  

It is convenient to rewrite $f_i$ as
\begin{equation}\label{f}
f_{i}=f_{i}^{(0)}+\Delta f_{i},
\end{equation}
where $f_i^{(0)}\equiv \mathrm{lim}_{k\to 0}f_i $. In \cite{our} the functions $\Delta f_i$ have been calculated in the framework of RPT, the result is 
\begin{eqnarray}\label{d_f}
\Delta f_{1} &=&\frac{F_{V}^{2}-2F_{V}G_{V}}{f_{\pi }^{2}}\biggl(\frac{1}{%
m_{\rho }^{2}}+\frac{1}{m_{\rho }^{2}- Q^2}\biggr)  \nonumber \\
&-&\frac{F_{A}^{2}}{f_{\pi }^{2}m_{a}^{2}}\biggl[ 2+\frac{(k\cdot l)^{2}}{%
D(l)D(-l)}+\frac{(Q^{2}+k\cdot Q)[4m_{a}^{2}-(Q^{2}+l^{2}+2k\cdot Q)] }{
8D(l)D(-l)}\biggr],  \label{eq:delta-f1} \\
\Delta f_{2} &=&-\frac{F_{A}^{2}}{f_{\pi }^{2}m_{a}^{2}}\frac{%
4m_{a}^{2}-(Q^{2}+l^{2}+2k\cdot Q)}{8D(l)D(-l)},  \label{eq:delta-f2} \\
\Delta f_{3} &=&\frac{F_{A}^{2}}{f_{\pi }^{2}m_{a}^{2}}\frac{k\cdot l}{%
2D(l)D(-l)}, 
\end{eqnarray}
for all notations and the details of calculation  see \cite{our}.  Taking the central value of the corresponding decay widths
\begin{eqnarray}
&&\Gamma(\rho^0\to e^+e^-)=6.85\pm0.11 \mathrm{keV}, \\
&&\Gamma(\rho^0\to\pi\pi)=150.7\pm2.9\mathrm{MeV}, \hspace{1.5em} \Gamma(a_1\to\pi\gamma)=640\pm240\mathrm{keV} \nonumber
\end{eqnarray}
and using the relations
\begin{eqnarray}
\Gamma(\rho^0\to\pi\pi)&=&\frac{G_V^2m_\rho^3}{48\pi f_\pi^4}(1-\frac{4m_\pi^2}{m_\rho^2})^{3/2} , \hspace{1.5em} \Gamma(\rho^0\to e^+e^-)=\frac{4\pi\alpha^2F_V^2}{3m_\rho} , \hspace{1.5em} \\
\Gamma(a_1\to\pi\gamma)&=&\frac{\alpha F_A^2 m_a}{24f_\pi^2}(1-\frac{m_\pi^2}{m_a^2})^3 , \nonumber
\end{eqnarray} 
we have the following values for the parameters of the model 
\begin{equation}
F_V=0.156 \mathrm{GeV}, \hspace{1em} G_V=0.066\mathrm{GeV}, \hspace{1em} F_A=0.122\mathrm{GeV},
\end{equation}
and $f_\pi=92.4$MeV.  In \cite{our} it was shown that the $\gamma^*\to\rho^\pm\pi^\mp\to\pi^+\pi^-\gamma$ is negligible and  we will discard it henceforward.

\textbf{3.} 
Our MC code for $e^+e^-\to\pi^+\pi^-\gamma$ is based on  the MC EVA structure \cite{binner}. 

The matrix element we use for the cross section in the MC simulation is
\begin{eqnarray}\label{cross_sect}
d\sigma&\sim&|M_{ISR}+M_{FSR}+M_\phi|^2 \\
&\simeq& |M_{ISR}+M_{FSR}|^2+|M_\phi|^2+2\mathrm{Re}(M_{FSR}^{(sQED)}\cdot M_\phi^*) , \nonumber
\end{eqnarray}
where we apply the EVA result for the initial state radiation (ISR) matrix element ($M_{ISR}$),  while the FSR matrix element ($M_{FSR}$) is taken from the RPT prediction ((\ref{fsr}) and (\ref{f}, \ref{d_f})) and $M_\phi$ is the amplitude for the $\phi$ direct decay. (We should mention here, that for initial state we will consider only the case of \textit{one} photon radiation, that corresponds to the LO approximation for the EVA generator.)  To estimate the $\phi\to\pi^+\pi^-\gamma$ decay we  apply the Achasov four quark parametrization \cite{achasov} with the parameters taken from the fit of the KLOE data for $\phi\to\pi^0\pi^0\gamma$  with only  the $f_0$ intermediate state \cite{kloepi0} (for  different parametrizations of the $\phi$ direct decay  and its contribution to the asymmetry and the cross section see \cite{czyz}, \cite{graz}). 

To begin with, we estimate the relative magnitude of the different contributions to cross section (\ref{cross_sect}) considering the inclusive kinematics:
\begin{eqnarray}    
0^\circ\leq\theta_\gamma\leq 180^\circ , \\
0^\circ\leq\theta_\pi\leq 180^\circ. \nonumber
\end{eqnarray} 
In the left part of Fig.1  the value of the different contributions to Eq. (\ref{cross_sect}) are shown for $s=m_\phi^2$. One can see that the $\phi$ resonant contribution (i.e. proportional to $|M_\phi|^2$) is quite large
and the additional RPT contribution to FSR (i.e. the contribution to FSR not included in the sQED$*$VMD model) can be revealed only in the case of  the destructive interference ($\mathrm{Re}(M_{FSR}^{sQED}\cdot  M_\phi^*)<0$). In this case the interference term and the $\phi$ resonant contribution almost cancel each other at the low $Q^2$ region.  The preliminary data  from the KLOE experiment  are in favour of this assumption \cite{kloe_phi}.

Within this assumption we consider the cuts used in the  KLOE large angle analysis \cite{kloe_large}:
\begin{eqnarray}\label{cuts}    
50^\circ\leq\theta_\gamma\leq 130^\circ , \\
50^\circ\leq\theta_\pi\leq 130^\circ. \nonumber
\end{eqnarray}

In Figs. 2 and 3 we show our numerical results for the cross section (\ref{cross_sect}) for  hard photon radiation with energies $\omega>20$ MeV. In Fig.3 the term $d\sigma_{sQED}$ corresponds to the right side of Eq. (\ref{cross_sect})  without the $\phi$-meson decay and FSR in sQED$*$VMD, the term $\sigma_{sQED+\phi}$ includes the $\phi$ contribution and $d\sigma_{TOT}$ is for the cross section (\ref{cross_sect}) with the $\phi$ term and FSR calculated by  RPT theory. As we can see, in the low $Q^2$ region, the additional FSR term is up to $30\%$ of the total contribution coming from  sQED and $\phi\to\pi^+\pi^-\gamma$
 decay.
In the left part of Fig.3 the peak about $1$GeV$^2$ corresponds to the $f_0$ intermediate state for the $\phi\to\pi\pi\gamma$ amplitude.

It has been proposed to take data outside the $\phi$ peak  ($s<m_\phi^2$), in order to reduce the background from  $\phi\to3\pi$ decay \cite{stef}. In this case the $\phi$ resonant contribution (the term $\sim |M_\phi|^2$ in Eq.(\ref{cross_sect})) is suppressed 
(see Fig.1, right) and 
and does not cancel
anymore with the interference term at low $Q^2$. The interference term however still
survives at $s=1$GeV$^2$, so that  
even for $s\lesssim m_\phi^2$ we cannot neglect the $\phi$ direct decay to the cross section (\ref{cross_sect}) (as it is shown in Fig. 4).  At the same time the value of the interference can be comparable with the FSR contribution  not covered by the sQED$*$VMD model. Therefore for a precise evaluation of the total contributions at low $Q^2$ the interference 
term and the contributions beyond sQED should be included. 


\textbf{4.} According to our numerical results the low $Q^2$ region is sensitive to the inclusion of additional FSR  contribution of RPT both for on-peak  and   off-peak energies.

Our results are limited by the main following reasons:
\begin{itemize}
\item we approximate the interference term by $M_{FSR}^{(sQED)}\cdot M_\phi^*$
\item 
in the the pion form-factor  in 
 RPT
 we  use only the $\rho$-meson contribution, whereas the actual
VMD results include also $\omega$ and $\rho'$
\item we do not include multiple photon emission (from initial and/or final 
state)
\item we parametrize the $\phi$ direct decay amplitude only through $f_0$ intermediate state .
\end{itemize} 


While the first three items can be improved in a  
refined version of the code, the precise description of 
the amplitude $\phi\to\pi\pi\gamma$ is a difficult task,
expecially at low $Q^2$. (We would like to stress again  that this energy region is of our interest because an essential FSR contribution beyond sQED$*$VMD can exist only for  the low $Q^2$ region, when the radiated photon is energetic.) 
The new data on $\phi\to\pi^0\pi^0\gamma$ from KLOE will certaintly 
help 
to refine the $\phi$ direct decay amplitude; 
some information 
can be extracted by a fit of the spectrum of $e^+e^-\to\pi^+\pi^-\gamma$ itself, using, for example, the charge asymmetry, as discussed in \cite{czyz}. 

Clearly, a model independent analysis of FSR contribution will be very useful. We hope to disantangle   FSR and the  $\phi$ direct decay contributions in a  model-independent way by using the information from cross section, asymmetry measurement for different beam energies.

 
Work is in progress.



\vspace{1cm}

\textbf{Acknowledgements} We thank F.Jegerlehner, J. K\"uhn, W. Kluge, H. Czy\' z and S. Eidelman for useful discussion. This work was supported in part through EU RTN Contract CT2002-0311.


\newpage


\begin{figure}[tbp]
\label{fig1}
\par
\parbox{1.05\textwidth}{
\includegraphics[width=0.45\textwidth,height=0.45\textwidth]{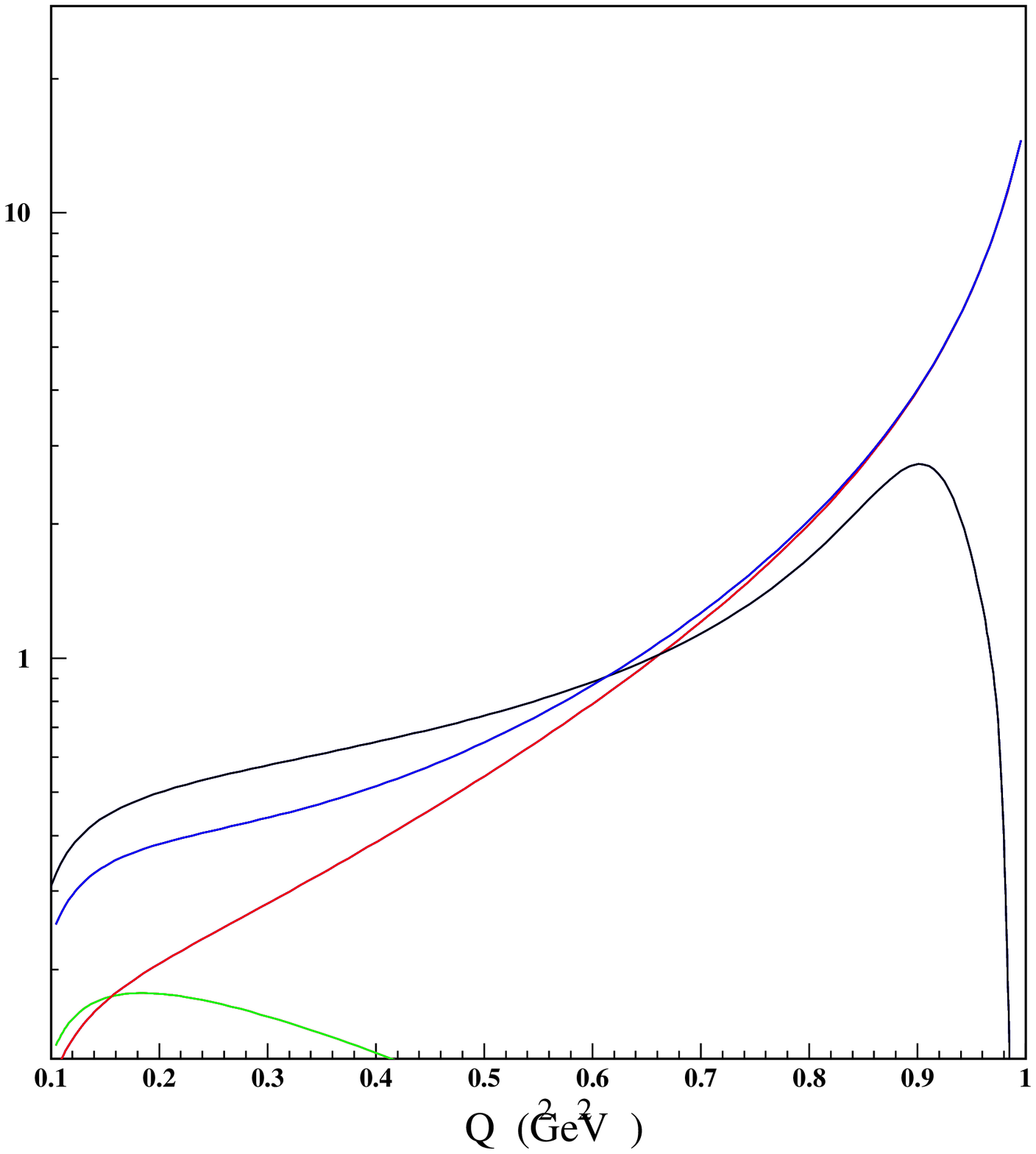}
\hspace{0.5cm}
\includegraphics[width=0.45\textwidth,height=0.45\textwidth]{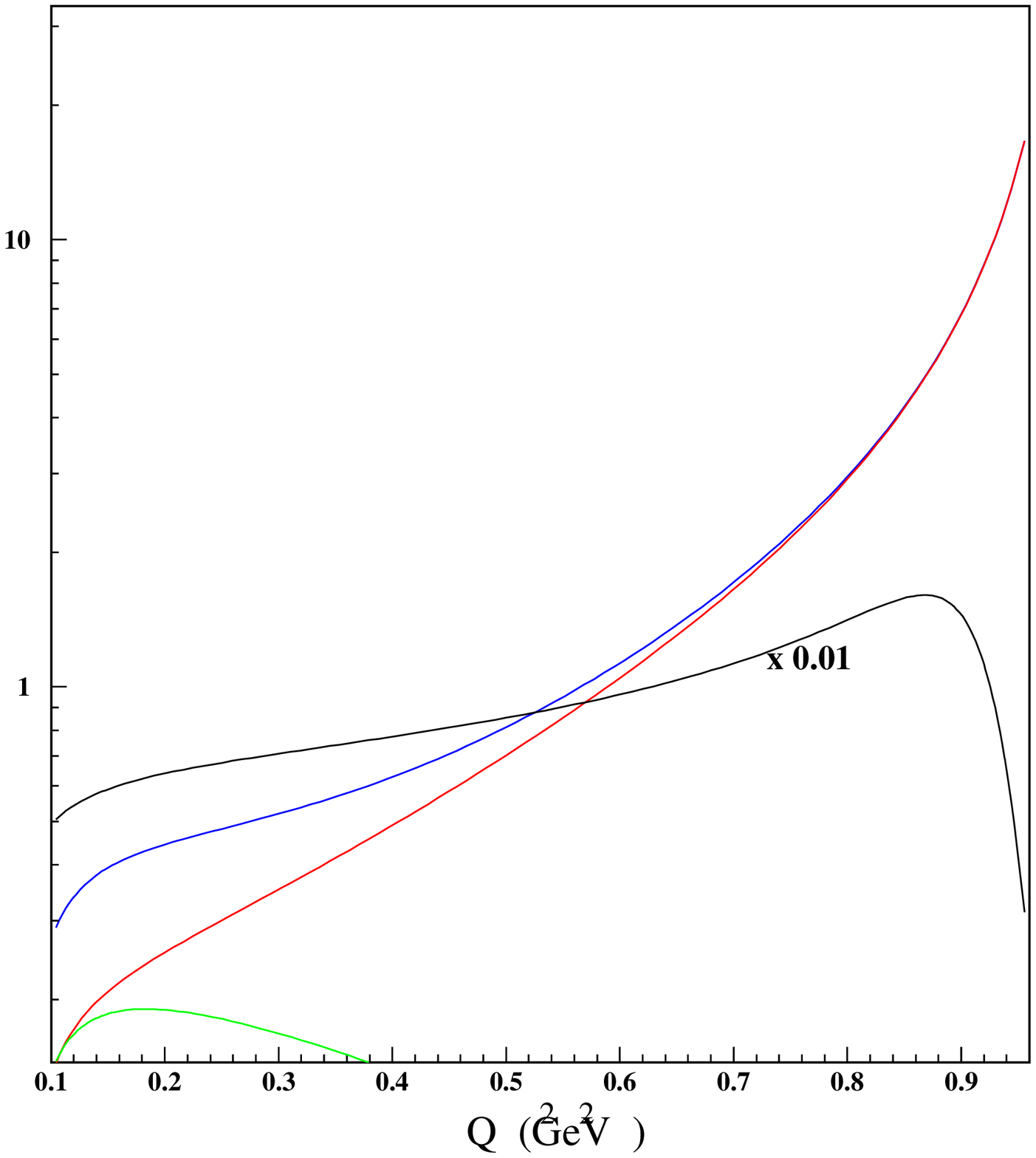}}
\caption{Contributions of FSR and $\phi$ direct decay to the cross section
$e^+e^-\to\pi^+\pi^- \gamma$, see Eq.~(\ref{cross_sect}). The black line corresponds to the $\phi$ resonant contribution. The blue line is the FSR contribution in the framework of RPT, the red line is FSR  in sQED, the green line is the additional RPT FSR contribution, beyond sQED. The left figure corresponds to $s=m_\phi^2$, the right one is for $s=1$ GeV$^2$ (i.e. below the $\phi$ resonance), where the $\phi$ resonant contribution is amplified by a factor $100$.}
\end{figure}
\begin{figure}[tbp]
\label{fig2}
\par
\parbox{1.05\textwidth}{\hspace{-0.4cm}
\includegraphics[width=0.5\textwidth,height=0.5\textwidth]{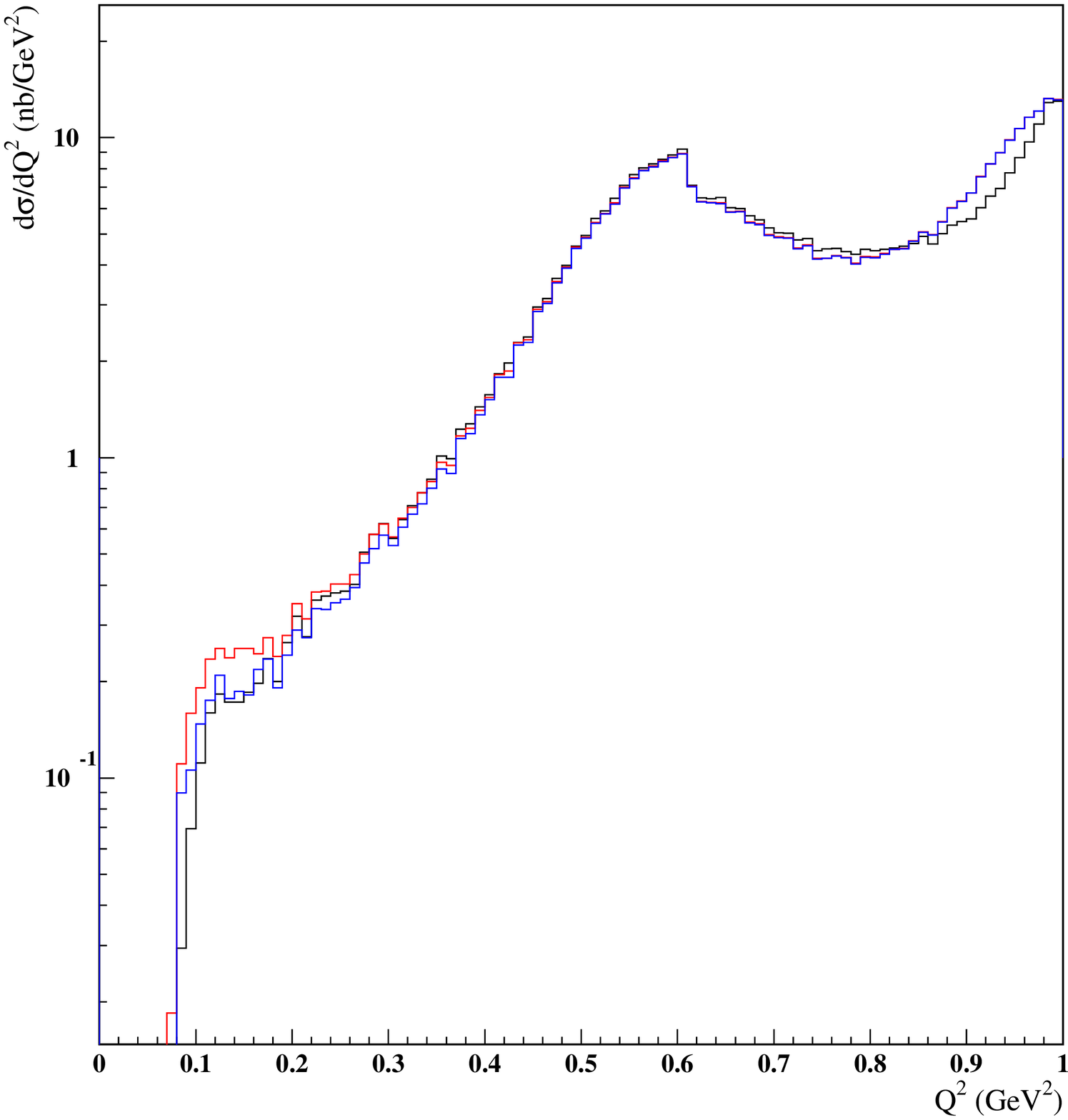}
\hspace{0.2cm}
\includegraphics[width=0.5\textwidth,height=0.5\textwidth]{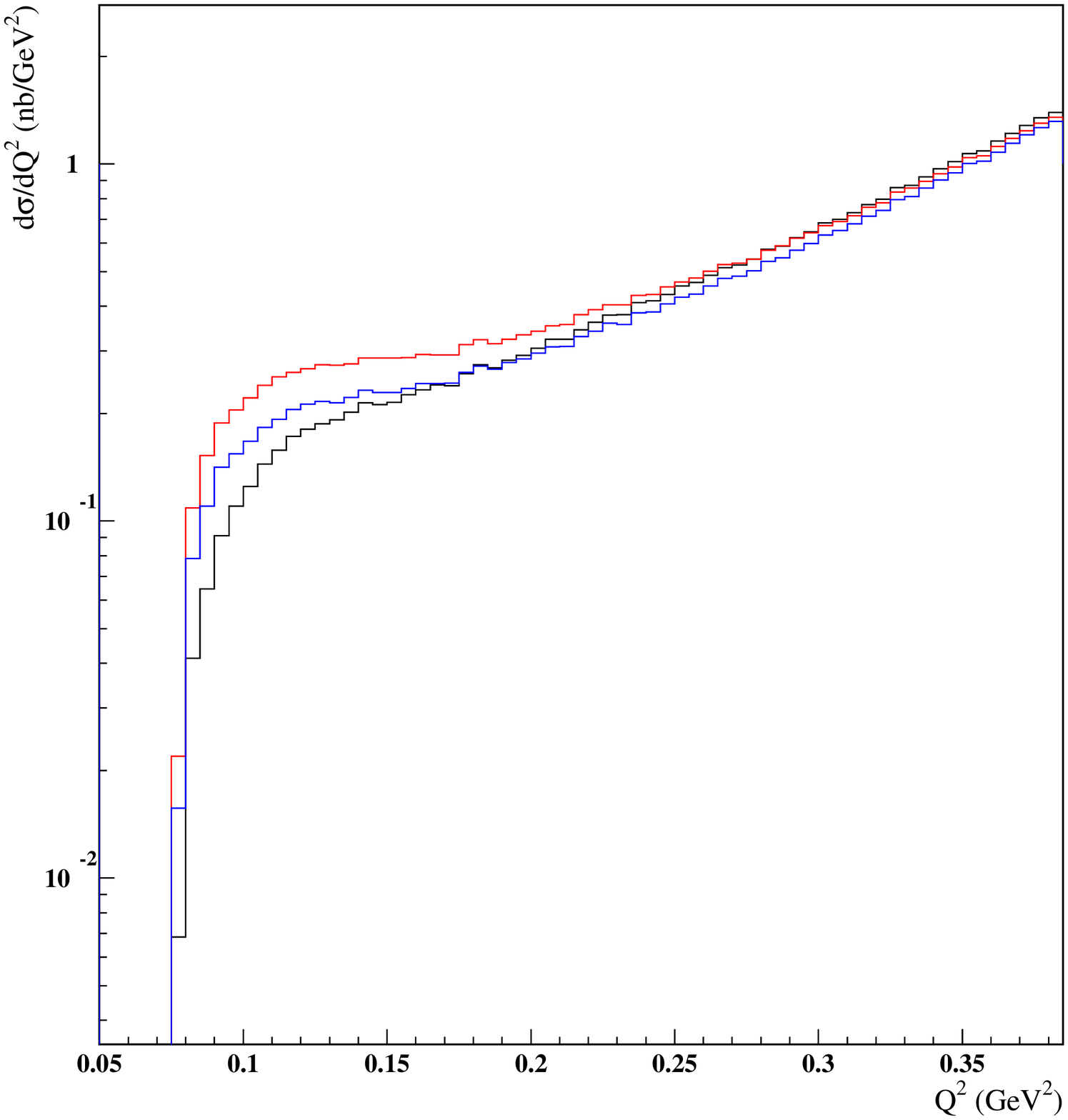}}
\caption{The  cross section
$e^+e^-\to\pi^+\pi^- \gamma$ with cuts (\ref{cuts}).
The black line corresponds to the result for Eq.~(\ref{cross_sect}) that includes FSR in the framework of the sQED$*$VMD model and  does not include the $\phi$ direct decay. The blue line is the same including  the $\phi$ decay. The red line corresponds to the cross section (\ref{cross_sect}) where FSR is calculated in RPT.  
The left figure is for the entire energy region, the right one is for the low $Q^2$ region.}
\end{figure}
\begin{figure}
\label{fig3}
\par
\parbox{1.01\textwidth}{\hspace{-0.3cm}
\includegraphics[width=0.5\textwidth,height=0.6\textwidth]{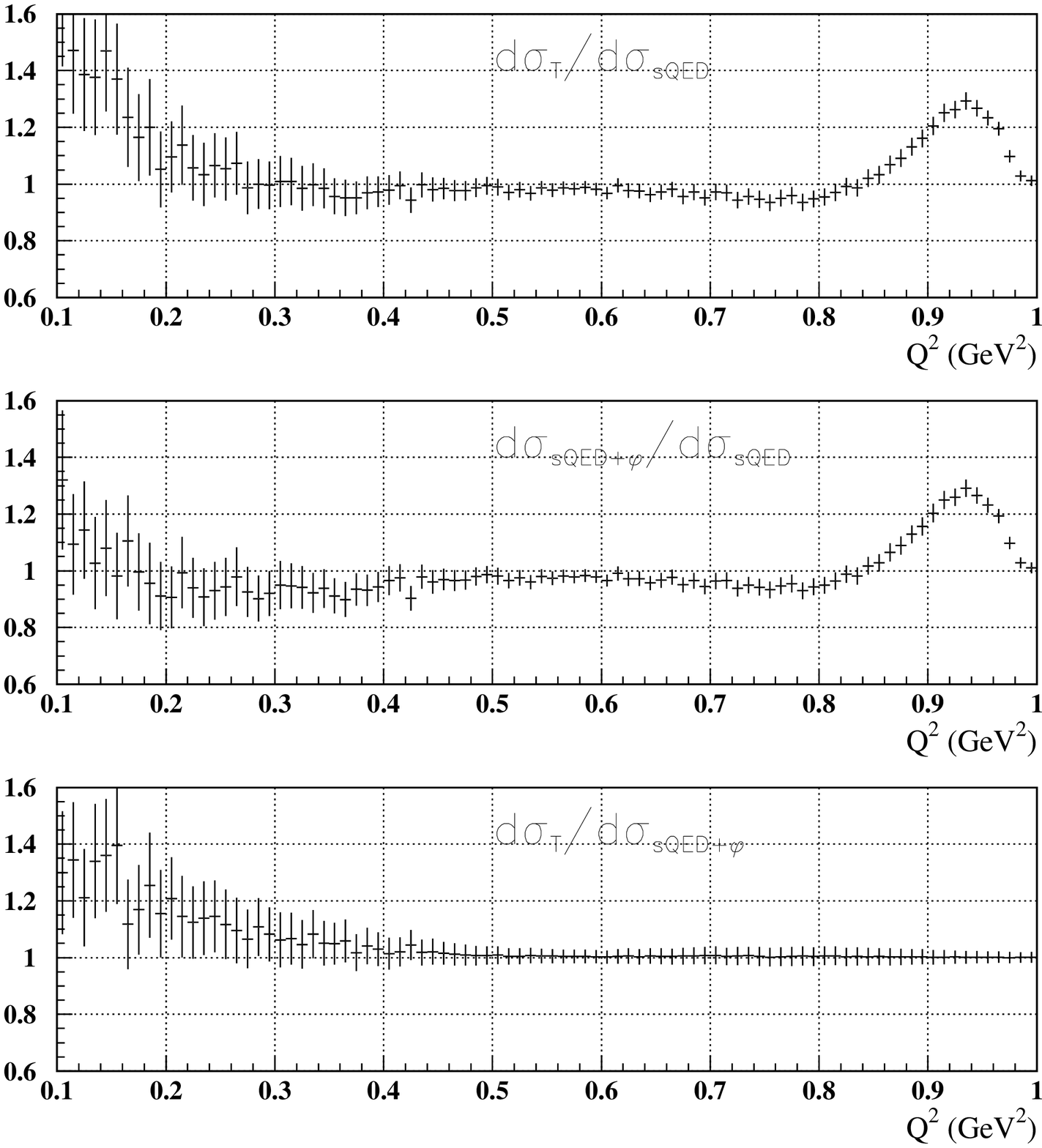}
\hspace{0.3cm}
\includegraphics[width=0.5\textwidth,height=0.6\textwidth]{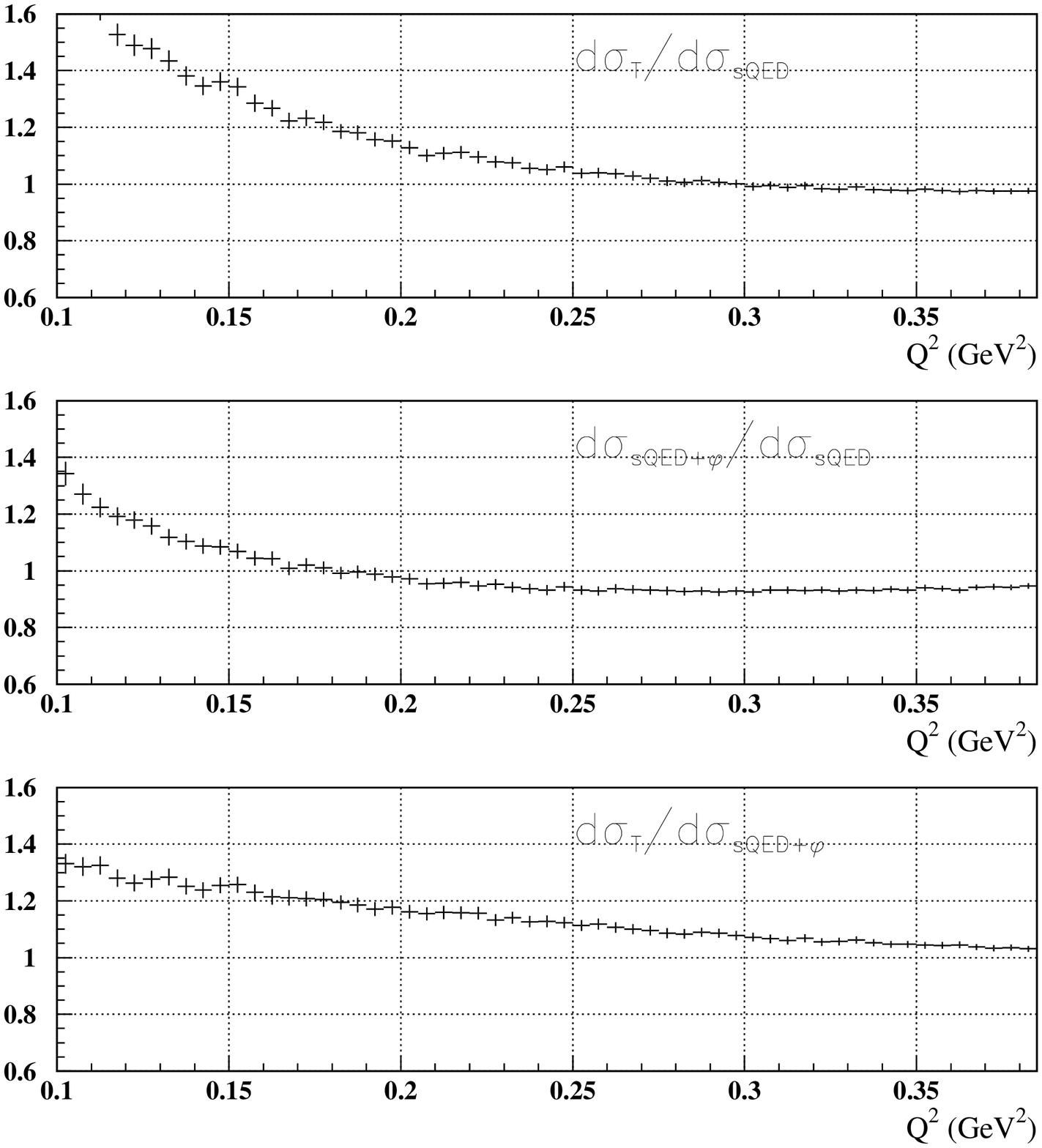}}
\vspace{-0.5cm} 
\caption{The relative value of the different contributions to the  cross section $e^+e^-\to\pi^+\pi^- \gamma$ process, see Eq.~(\ref{cross_sect}), for $s=m_\phi^2$ and with cuts (\ref{cuts}).  The left figure  corresponds to the entire $Q^2$ region, the right one is for the low $Q^2$ region.}
\end{figure}
\vspace{-1cm}
\begin{figure}
\label{fig4}
\par
\parbox{1.01\textwidth}{\hspace{-0.3cm}
\includegraphics[width=0.5\textwidth,height=0.6\textwidth]{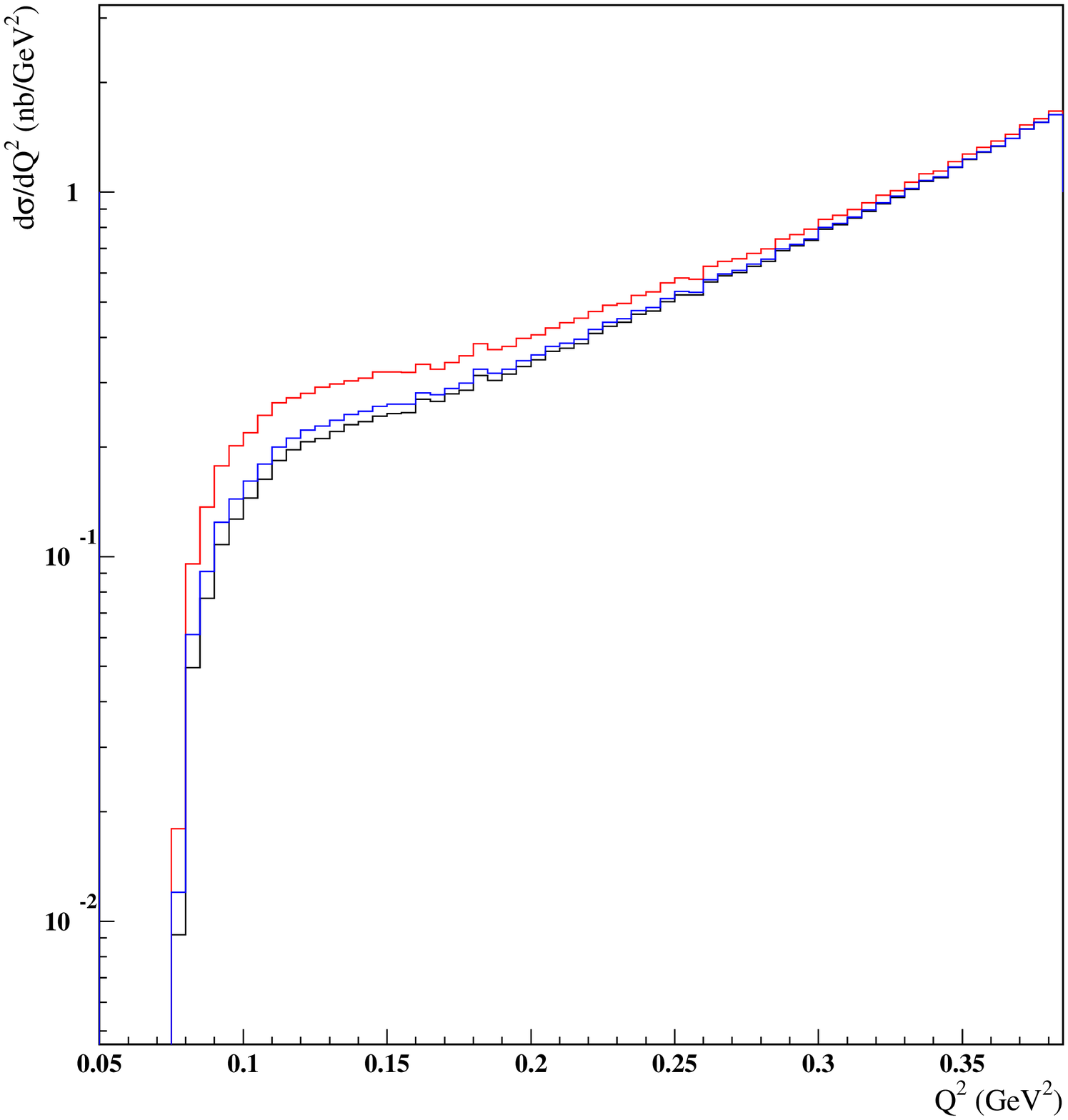}
\hspace{0.3cm}
\includegraphics[width=0.5\textwidth,height=0.6\textwidth]{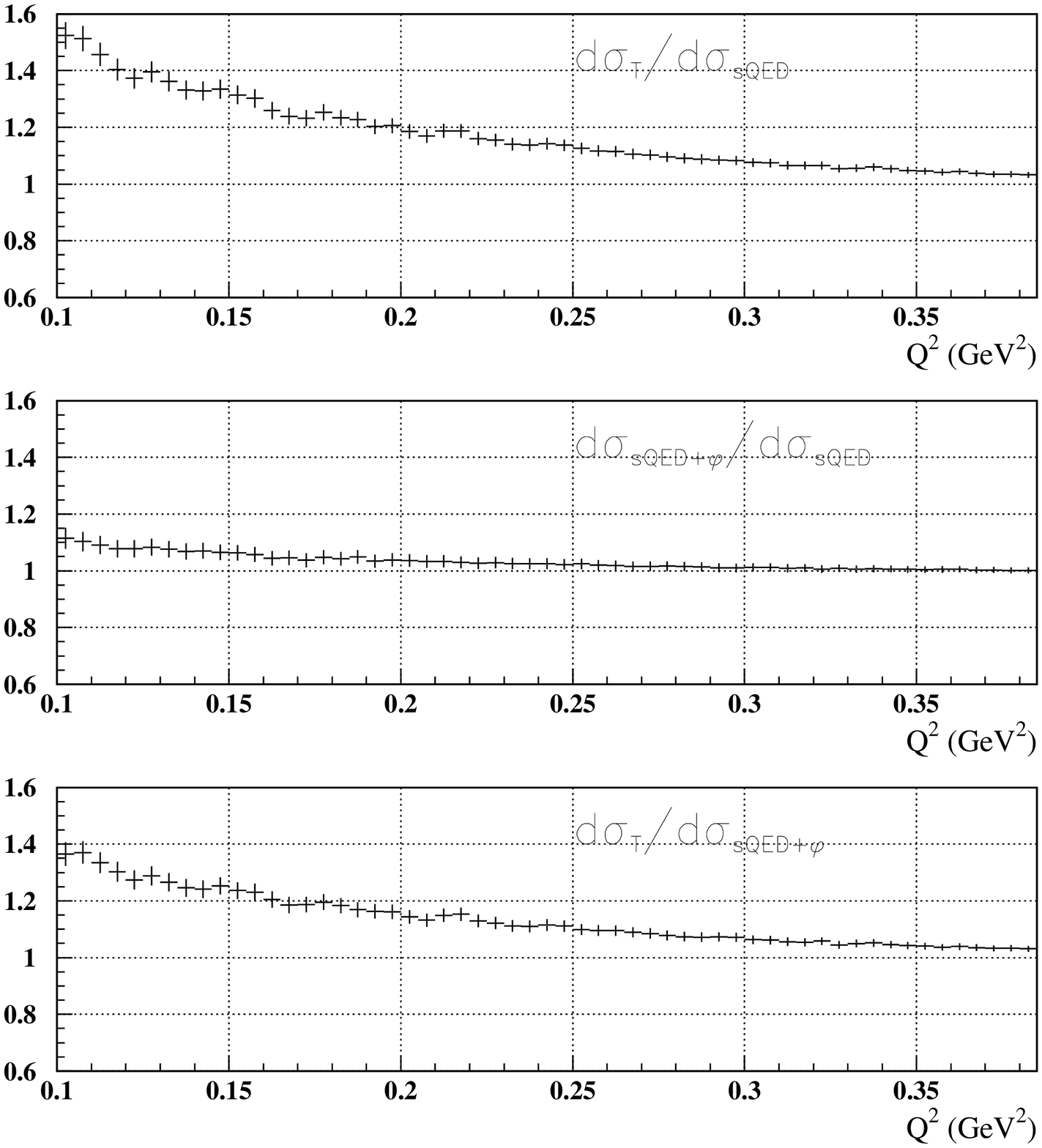}}
\vspace{-1.5cm} 
\caption{The  cross section
$e^+e^-\to\pi^+\pi^- \gamma$ process, see (\ref{cross_sect}), for $s=1$GeV$^2$, cuts (\ref{cuts}) and low $Q^2$. Left: the absolute value of the cross section (notations for the curves the same as in Fig.2). Right: the relative value of the different contributions to the cross section (\ref{cross_sect}).}
\end{figure}

\end{document}